\newcommand{\be}{\begin{equation}}\newcommand{\ee}{\end{equation}}
\newcommand{\bea}{\begin{eqnarray}}\newcommand{\eea}{\end{eqnarray}}
\newcommand{\nn}{\nonumber\\}\newcommand{\p}[1]{(\ref{#1})}
\documentstyle[12pt]{article}
\begin{document}
\begin{titlepage}
\begin{flushright}
LNF-00/014 (P) \\
LPTHE 00-15\\
JINR-E2-2000-68\\
hep-th/0003273\\
March 2000
\end{flushright}
\vskip 0.6truecm
\begin{center}{\Large\bf
Superworldvolume dynamics of superbranes from
nonlinear  realizations}
\end{center}
 \vskip 0.6truecm
\centerline{\bf S. Bellucci${}^{\, a,1}$, E. Ivanov${}^{\,b,c,2}$, S.
Krivonos${}^{\,c,3}$}  \vskip 0.6truecm

\centerline{$^a${\it INFN, Laboratori Nazionali di Frascati, P.O. Box
13, I-00044 Frascati, Italy}}

\vspace{0.2cm}
\centerline{$^b$ {\it
Laboratoire de Physique Th\'eorique et des Hautes Energies,}}
\centerline{\it Unit\'e associ\'ee au CNRS UMR 7589, Universit\'e
Paris 7,}
\centerline{\it  2 Place Jussieu, 75251 Paris Cedex 05, France}

\vspace{0.2cm}
\centerline{$^c${\it Bogoliubov Laboratory of
Theoretical Physics, JINR,}}
\centerline{\it 141 980 Dubna, Moscow region,
Russian Federation}
\vskip 0.6truecm  \nopagebreak

\begin{abstract}
\noindent Based on the concept of the partial
breaking of global supersymmetry (PBGS), we derive the
worldvolume superfield equations of motion for $N=1, \,D=4$
supermembrane, as well as for the space-time filling D2- and
D3-branes, from nonlinear realizations of
the corresponding supersymmetries. We argue that it is of no need
to take care of the relevant automorphism groups when being
interested in the dynamical equations. This essentially
facilitates computations.
As a by-product, we obtain a new
polynomial  representation for the $d=3,4$ Born-Infeld equations,
with merely a cubic nonlinearity.

\end{abstract}
\vspace{0.5cm}

\begin{flushleft}
\hspace*{0.91cm} PACS 11.17.+y, 11.30.Pb
\end{flushleft}
\vfill

\noindent{\it E-Mail:}\\
{\it 1) bellucci@lnf.infn.it}\\
{\it 2) eivanov@lpthe.jussieu.fr,  eivanov@thsun1.jinr.ru}\\
{\it 3) krivonos@thsun1.jinr.ru}
\newpage

\end{titlepage}

\noindent{\bf 1. Introduction.} During last few years there was a considerable
interest in applying the general method of nonlinear
realizations to systems with partial breaking of global
supersymmetries (PBGS), first of all to the superbranes as a
notable  example of such systems (see, e.g., \cite{BIK1,BIK2,PW} and
refs. therein). On this path one meets two problems. The first one is purely
computational. Following the general prescriptions of nonlinear
realizations, one is led to include into the coset, alongside with the
spontaneously broken translation and supertranslation generators, also the
appropriate part of generators of the automorphism group for the given
supersymmetry algebra (including those of the Lorentz group). This makes the
computations beyond the linearized approximation rather complicated. Moreover,
sometimes these additional symmetries which we should take into account at the
step of doing the coset routine  appear to be explicitly broken at the level
of the invariant action (see, e.g., refs. \cite{BG0,BG3,rt}), with no clear
reasons for this. The second, closely related difficulty is lacking of a
systematic procedure for constructing the PBGS actions. In all the cases
elaborated so far, the PBGS Lagrangians cannot be constructed in a manifestly
invariant way from the relevant Cartan forms: under the broken supersymmetry
transformations they are shifted  by the spinor or $x$-derivatives (like the
WZNW or Chern-Simons Lagrangians).

In the present note we argue, on several instructive examples, that the
automorphism symmetries can be ignored if we are interested only in the
equations of motion for the given PBGS system. This radically
simplifies the  calculations,
resulting in rather simple manifestly covariant equations in which all
nonlinearities are hidden inside the covariant derivatives.

\vspace{0.3cm}
\noindent{\bf 2. $N=1, \, D=4$ supermembrane and D2-brane.} To
clarify the main idea of our approach, let us start from the well known
systems with partially broken global supersymmetries \cite{AGIT,IK1}.
Our goal is to get the corresponding superfield equations of motion in terms of
the worldvolume superfields starting from the nonlinear realization of the
global supersymmetry group.

The supermembrane in $D=4$ spontaneously breaks half of four $N=1,D=4$
supersymmetries
and one translation. Let us split the set of generators of $N=1\; D=4$
Poincar\'e superalgebra
(in the $d=3$ notation) into the unbroken $\left\{ Q_a, P_{ab} \right\}$ and
broken $\left\{ S_a, Z
\right\}$ ones ($a,b=1,2$). The $d=3$ translation generator $P_{ab} =
P_{ba}$ together
with the generator $Z$ form the $D=4$ translation generator.
The basic anticommutation relations read \footnote{Hereafter, we consider the
spontaneously broken supersymmetry algebras modulo possible extra
central-charge type terms which should be present in the full algebra of the
corresponding Noether currents to evade the no-go theorem of ref.
\cite{Wit} along the lines of ref. \cite{hlp}.}
\be \left\{
Q_{a},Q_{b}\right\}=P_{ab}\; ,\quad \left\{ Q_{a},S_{b}\right\} =
\epsilon_{ab}Z\; , \quad \left\{ S_{a},S_{b}\right\} = P_{ab} \;. \label{susy}
\ee

In contrast to our previous considerations \cite{IK1,BIK1,BIK2}, here
we prefer
to construct the nonlinear realization of the superalgebra \p{susy}
itself, ignoring all  generators of the automorphisms of \p{susy} (the
spontaneously broken as well as unbroken ones), including those of
$D=4$ Lorentz group $SO(1,3)$. Thus, we
put all generators  into the coset and associate
the $N=1\,,\, d=3$ superspace coordinates $\left\{
\theta^a, x^{ab} \right\}$ with $Q_a, P_{ab}$. The remaining coset
parameters are
Goldstone superfields, $\psi^a \equiv \psi^a(x,\theta),\;q \equiv
q(x,\theta)$. A coset element $g$ is defined by
\be\label{coset}
g=e^{x^{ab}P_{ab}}e^{\theta^{a}Q_{a}}e^{qZ}
  e^{\psi^aS_a} \;.
\ee
As the next step of the coset formalism, one constructs the
Cartan 1-forms
\be
g^{-1}d g =  \omega_Q^aQ_a + \omega_P^{ab} P_{ab} + \omega_ZZ +
\omega_S^a S_a , \label{cartan1}
\ee
\bea
\omega_Z & = &  dq+\psi_{a}d\theta^{a}\; , \;
\omega_P^{ab} =dx^{ab}+\frac{1}{4}\theta^{(a}d\theta^{b)} +
  \frac{1}{4}\psi^{(a} d\psi^{b)}  \; ,\nn
\omega_Q^a &=&  d\theta^{a} \; ,\; \omega_S^a=d\psi^{a};.
\label{cartan}
\eea
and define the covariant derivatives
\be\label{cd}
{\cal D}_{ab} =  (E^{-1})^{cd}_{ab}\,\partial_{cd} \; , \quad
{\cal D}_a = D_a + \frac{1}{2}\psi^b D_a \psi^c \,{\cal D}_{bc} =
         D_a + \frac{1}{2}\psi^b {\cal D}_a \psi^c \,{\partial}_{bc}~,
\ee
where
\bea
&&D_a=\frac{\partial}{\partial \theta^a}+
\frac{1}{2}\theta^b\partial_{ab}\; , \quad
\left\{ D_a, D_b \right\} =\partial_{ab} \; , \label{flatd} \\
&& E_{ab}^{cd}=\frac{1}{2}(\delta_a^c\delta_b^d+\delta_a^d\delta_b^c)+
  \frac{1}{4}(\psi^c\partial_{ab}\psi^d+ \psi^d\partial_{ab}\psi^c) \;.
\eea
They obey the following algebra
\bea
&&\left[ {\cal D}_{ab},{\cal D}_{cd} \right] =
-{\cal D}_{ab}\psi^f{\cal D}_{cd}\psi^g \,
           {\cal D}_{fg} \; , \nn
&&\left[ {\cal D}_{ab},{\cal D}_{c} \right] =
{\cal D}_{ab}\psi^f{\cal D}_{c}\psi^g\,
           {\cal D}_{fg} \; , \nn
&&\left\{ {\cal D}_{a},{\cal D}_{b} \right\} ={\cal D}_{ab}+
        {\cal D}_{a}\psi^f{\cal D}_{b}\psi^g\,
           {\cal D}_{fg} \; .  \label{algebra}
\eea
Not all of the above Goldstone superfields
$\left\{ q(x,\theta),\psi^a(x,\theta)\right\}$
must be treated as independent.
Indeed,  $\psi_{a}$ appears
inside the form $\omega_Z$  {\it linearly} and so can be covariantly
eliminated by the manifestly covariant constraint (inverse Higgs effect
\cite{invh})
\be
\left. \omega_Z\right|_{d\theta} = 0 \Rightarrow \psi_a={\cal D}_a q \;,
\label{basconstr}
\ee
where $|_{d\theta}$ means the ordinary $d\theta$-projection of the form.
Thus the superfield
$q(x,\theta)$ is the only essential Goldstone superfield needed to
present the partial spontaneous breaking $N=1\,,\; D=4
\;\Rightarrow \; N=1\,,\; d=3$ within the coset scheme.

Now we are ready to put additional, manifestly covariant constraints on
the
superfield $q(x,\theta)$, in order to get dynamical equations. The main
idea is to
covariantize the ``flat'' equations of motion. Namely, we simply replace
the flat covariant derivatives in the standard equation of motion for
the
bosonic scalar superfield in $d=3$
\be\label{flateom1}
D^a D_a q=0
\ee
by the covariant ones \p{cd}
\be\label{eom1}
{\cal D}^a {\cal D}_a q=0 \;.
\ee
The equation \p{eom1} coincides with the equation of motion of the
supermembrane in $D=4$ as it was presented in \cite{IK1}. Thus, we conclude
that, at least in this specific case, additional superfields-parameters of the
extended coset with all the automorphism symmetry generators included are
auxiliary and can be dropped out if we are interested in the equations of
motion only.

Actually, in \cite{IK1}
eq. \p{eom1} was deduced, proceeding from the $D=4$ Lorentz covariant coset
formalism  with preserving all initial symmetries. This means that
\p{eom1}, having been now reproduced from the coset involving only the
translations and supertranslations generators, possesses the hidden
covariance under the full $D=4$ Lorentz group. On the other hand, one
more automorphism symmetry of the $N=1,\, D=4$ supersymmetry algebra,
``$\gamma_5$'' symmetry, is explicitly
broken in eq. \p{eom1}, and there is no way to keep it. In the $d=3$
notation this symmetry is realized as an extra $SO(2)$ with respect
to which the generators $Q_a$ and $S_b$ and, respectively, the coset
parameters $\theta^a, \psi^a$ form a 2-vector. This
symmetry is spontaneously broken at the level of the
transformation laws, with the auxiliary field
of $q(x,\theta)$ being the relevant Goldstone field. From eq. \p{eom1} we
conclude that it  cannot be preserved even in this spontaneously broken
form when $q$ is subjected to the dynamical equation: one can preserve
the  spontaneously broken $D=4$ Lorentz symmetry at most. This $U(1)$ is
explicitly  broken in the off-shell PBGS  action of ref. \cite{IK1}, as well as
in the corresponding Green-Schwarz  action \cite{AGIT}. A similar
phenomenon  was observed in refs. \cite{BG0,BG3} for the $N=(1,0), \,D=6$
3-brane. There, the auxiliary fields of
the basic worldvolume $N=1, \,d=4$ Goldstone  chiral supermultiplet
are the Goldstone fields  parameterizing the coset $SU(2)_A/U(1)_A$ of
the automorphism  $SU(2)_A$ group of $N=(1,0), \,D=6$ Poincar\'e
superalgebra, and the coset part of $SU(2)_A$ is realized
as nonlinear shifts of these fields. In the
superfield equations of  motion of the 3-brane and the corresponding off-shell
action  this $SU(2)_A$ is {\it explicitly} broken down to $U(1)_A$,
though the spontaneously broken $D=6$ Lorentz symmetry is still preserved.

As a straightforward application of the idea that the automorphism
symmetries are irrelevant when deducing the equations of motion,
let us consider the case of the ``space-time filling'' D2-brane (i.e. having
no scalar fields in its worldvolume multiplet the field content of which
is that of $N=1,\,d=3$ vector multiplet). The
main problem with the description of D-branes within the standard nonlinear
realization approach is the lack of the coset generators to which one could
relate the gauge fields as the coset parameters \footnote{For the covariant
field strengths as Goldstone fields such generators can still be found in the
automorphism symmetry algebras \cite{PW,EI}.}. So we do not know how interpret
the gauge fields as coset parameters in this case \footnote{It seems that the
existing interpretation of gauge fields as the coset fields \cite{IO} can be
generalized to the PBGS case only on the way of non-trivial unification of the
gauge group algebra with that of supersymmetry, so that the gauge group
transformations appeared in the closure of supersymmetries before any
gauge-fixing as a sort of tensorial central charges.}.  Let us show how these
difficulties can be circumvented in the present approach.

The superalgebra we start with is the same algebra \p{susy}, but
now without the central charge
$$ Z=0 \;.$$
The coset element $g$ contains only one Goldstone superfield $\psi^a$
which now must be treated as the essential one, and the covariant derivatives
coincide
with \p{cd}. Bearing in mind to end up with the irreducible field
content of $N=1,\,d=3$ vector multiplet, we are led to treat
$\psi^a$ as the corresponding  superfield strength and to find the
appropriate covariantization of the  flat irreducibility constraint and the
equation of motion. In the flat case the $d=3$ vector multiplet is represented
by a $N=1$ spinor superfield strength $\mu_a$ subjected to the Bianchi identity
\cite{bibl}: \be\label{cc1} D^a\mu_a=0 \; \Rightarrow \; \left\{
 \begin{array}{l}
   D^2 \mu_a=-\partial_{ab}\mu^b~,  \nn
   \partial_{ab}D^a\mu^b = 0~ . \nonumber
  \end{array} \right\}.
\ee
This leaves in $\mu_a$ the first fermionic
(Goldstone) component, together with the
divergenceless vector $F_{ab}\equiv D_a\mu_b|_{\theta=0}$
(i.e., just the gauge field strength). The equation of motion reads
\be\label{flateom2}
D^2 \mu_a =0 \; .
\ee
In accordance with our approach, we propose the following equations
which should describe the D2-brane:
\be\label{eom2}
 (a)\;\;{\cal D}^a\psi_a=0~, \quad (b)\; \; {\cal D}^2 \psi_a =0~.
\ee
The equation $(a)$ is a covariantization of the irreducibility
constraint \p{cc1} while $(b)$ is the covariant equation of motion.

In order to see which kind of dynamics is encoded in \p{eom2}, we
considered it
in the bosonic limit. We found that it amounts to the following
equations
for the vector $V_{ab}\equiv {\cal D}_a\psi_b|_{\theta=0}$:
\be\label{boseq1}
\left( \partial_{ac} +V_a^m V_c^n \partial_{mn}\right) V_b^c =0 \;.
\ee
One can wonder how these nonlinear but polynomial equations can be related
to the nonpolynomial  Born-Infeld theory which is just the bosonic core of
the superfield D2-brane theory as was explicitly demonstrated in \cite{IK1}.
The trick is to rewrite the parts of the equation \p{boseq1}, respectively
antisymmetric and symmetric in the indices $\left\{ a,b \right\}$, as follows:
\bea
&& \partial_{ab} \left( \frac{V^{ab}}{2-V^2} \right)=0 \;,
\label{boseq2}\\
&& \partial_{ac} \left( \frac{ V_b^c}{2+V^2}\right)+
 \partial_{bc} \left( \frac{ V_a^c}{2+V^2}\right) =0 \;, \label{boseq3}
\eea
where $V^2\equiv V^{mn}V_{mn}$.
After passing to the ``genuine'' field strength
\be\label{fs}
F^{ab}=\frac{2V^{ab}}{2-V^2} \Rightarrow
\partial_{ab}F^{ab}=0\;,
\ee
the equation of motion \p{boseq3} takes the familiar Born-Infeld form
\be
\partial_{ac} \left( \frac{ F_b^c}{\sqrt{1+2F^2}}\right)+
 \partial_{bc} \left( \frac{ F_a^c}{\sqrt{1+2F^2}}\right) =0 \;.
\label{boseq4}
\ee
Thus we have proved that the bosonic part of our system \p{eom2} indeed
coincides with the Born-Infeld equations. One may explicitly show that the full
equations \p{eom2} are equivalent to the worldvolume superfield equation
following from the off-shell D2-brane action given in \cite{IK1} (augmented
with the Bianchi identity \p{cc1}). An indirect proof is based on the fact
that \p{eom2} is an $N=1$ extension of the bosonic $d=3$
Born-Infeld equations, such that it possesses one more nonlinearly realized
supersymmetry completing the explicit one to $N=2, \, d=3$ superalgebra
\p{susy} with $Z=0$. On the other hand, the $N=1, \,d=3$ superfield action of
\cite{IK1}  is uniquely specified by requiring it to possess this second
supersymmetry.  Hence both types of equations should be equivalent.

In closing this Section, it is worth mentioning that the equations \p{boseq1}
which equivalently describe the bosonic Born-Infeld dynamics in $d=3$, look
much simpler than the standard ones \p{fs}, \p{boseq4}.

\vspace{0.3cm}

\noindent{\bf 3. D3-brane.} As another interesting application of the
proposed approach, we shall consider the space-time filling D3-brane in $d=4$.
This system amounts to the PBGS pattern $N=2,\,d=4\;\rightarrow \;N=1,\,d=4$,
with a nonlinear generalization of $N=1,\,d=4$ vector multiplet as the
Goldstone multiplet \cite{BG2,rt}. The off-shell superfield action for this
system and the related equations of motion are known \cite{BG2}, but the
latter have never been derived directly from the coset approach.

Our starting point is the $N=2,\, d=4$ Poincar\'e superalgebra {\it
without}
central charges:
\be
\left\{ Q_{\alpha}, {\bar Q}_{\dot\alpha} \right\}=2P_{\alpha\dot\alpha}
\;,\;
\left\{ S_{\alpha}, {\bar S}_{\dot\alpha} \right\}=2P_{\alpha\dot\alpha}
\;.\label{susyD4}
\ee
Assuming the $S_{\alpha}, {\bar S}_{\dot\alpha}$ supersymmetries
to be spontaneously broken, we introduce the Goldstone superfields
$\psi^{\alpha}(x,\theta,\bar\theta), \,
{\bar\psi}^{\dot\alpha}(x,\theta,\bar\theta)$ as the corresponding parameters
in the following coset (we use the same notation as in \cite{BG2})
\be
g=e^{ix^{\alpha\dot\alpha}P_{\alpha\dot\alpha}}
   e^{ i\theta^{\alpha}Q_{\alpha}+
   i{\bar\theta}_{\dot\alpha}{\bar Q}^{\dot\alpha}}
e^{ i\psi^{\alpha}S_{\alpha}+
   i{\bar\psi}_{\dot\alpha}{\bar S}^{\dot\alpha}} \;.
\ee
With the help of the Cartan forms
\bea\label{cf5}
g^{-1}dg & =& i\omega^{\alpha\dot\alpha}P_{\alpha\dot\alpha}+
  i\omega_Q^{\alpha}Q_{\alpha}+
  i{\bar\omega}_{Q\;\dot\alpha}{\bar Q}^{\dot\alpha}+
  i\omega_S^{\alpha}S_{\alpha}+
  i{\bar\omega}_{S\;\dot\alpha}{\bar S}^{\dot\alpha}\;,\nn
\omega^{\alpha\dot\alpha} & = & dx^{\alpha\dot\alpha}-i\left(
  \theta^{\alpha}d{\bar\theta}^{\dot\alpha}+
  {\bar\theta}^{\dot\alpha}d\theta^{\alpha}+
\psi^{\alpha}d{\bar\psi}^{\dot\alpha}+
  {\bar\psi}^{\dot\alpha}d\psi^{\alpha}\right) \;, \nn
\omega_Q^{\alpha}&=&d\theta^{\alpha}\;,\;
{\bar\omega}_Q^{\dot\alpha}=d{\bar\theta}^{\dot\alpha}\;,\;
\omega_S^{\alpha}=d\psi^{\alpha}\;,\;
{\bar\omega}_S^{\dot\alpha}=d{\bar\psi}^{\dot\alpha}\;,
\eea
one can define the covariant derivatives
\bea\label{cd4}
{\cal D}_{\alpha\dot\alpha}&=&
 \left(
E^{-1}\right)_{\alpha\dot\alpha}^{\beta\dot\beta}\partial_{\beta\dot\beta}
  \;, \nn
{\cal D}_{\alpha}&=& D_{\alpha}-i\left(
{\bar\psi}^{\dot\beta}D_{\alpha}\psi^{\beta} +
  \psi^{\beta}D_{\alpha}{\bar\psi}^{\dot\beta}\right) {\cal
   D}_{\beta\dot\beta} \; , \nn
{\overline{\cal D}}_{\dot\alpha}&=& {\bar D}_{\dot\alpha}-
i\left( {\bar\psi}^{\dot\beta}{\bar D}_{\dot\alpha}\psi^{\beta} +
  \psi^{\beta}{\bar D}_{\dot\alpha}{\bar\psi}^{\dot\beta}\right) {\cal
   D}_{\beta\dot\beta} \; ,
\eea
where
\be
E_{\alpha\dot\alpha}^{\beta\dot\beta}= \delta_{\alpha}^{\beta}
 \delta_{\dot\alpha}^{\dot\beta}
-i\psi^{\beta}\partial_{\alpha\dot\alpha}{\bar\psi}^{\dot\beta}-
i{\bar\psi}^{\dot\beta}\partial_{\alpha\dot\alpha}\psi^{\beta}\;,
\ee
and the flat covariant derivatives are defined as follows
\be
D_{\alpha}=\frac{\partial}{\partial\theta^{\alpha}} -
i{\bar\theta}^{\dot\alpha}\partial_{\alpha\dot\alpha}~, \quad
{\bar D}_{\dot\alpha}=
 -\frac{\partial}{\partial{\bar\theta}^{\dot\alpha}} +
i{\theta}^{\alpha}\partial_{\alpha\dot\alpha}\;.
\ee
Now we are ready to write the covariant version of the constraints
on $\psi^{\alpha},\,{\bar\psi}^{\dot\alpha}$ which define the superbrane
generalization of $N=1,\,d=4$ vector multiplet, together with the covariant
equations of motion for this system.

As is well-known \cite{bw}, the $N=1,\,d=4$ vector multiplet is
described by
a chiral $N=1$ field strength $W_\alpha \,$,
\be\label{chir5}
{\overline D}_{\dot\alpha}W_{\alpha}=0~, \quad
D_{\alpha}{\overline W}_{\dot\alpha}=0~, \ee
which satisfies the irreducibility constraint (Bianchi identity)
\be\label{constr5}
D^{\alpha}W_{\alpha}+{\overline D}_{\dot\alpha}{\overline
W}^{\dot\alpha}=0
\;.
\ee
The free equations of motion for the vector multiplet read
\be\label{eom5}
D^{\alpha}W_{\alpha}-{\overline D}_{\dot\alpha}{\overline
W}^{\dot\alpha}=0
\;.
\ee

It was shown in \cite{BG2} that the chirality constraints \p{chir5} can
be directly covariantized
\be\label{chir5e}
{\overline{\cal D}}_{\dot\alpha}\psi_{\alpha}=0~, \quad {\cal
D}_{\alpha}\bar\psi_{\dot\alpha}=0~.  \ee
These conditions are compatible with the algebra of the
covariant  derivatives \p{cd4}. This algebra, with the constraints
\p{chir5e} taken into
account, reads
\cite{BG2} \bea && \{{\cal D}_\alpha, \,{\cal D}_\beta \} =
\{{\overline{\cal D}}_{\dot\alpha}, \, {\overline{\cal D}}_{\dot\beta} \} =
0~, \nn && \{{\cal D}_\alpha, \,{\overline{\cal D}}_{\dot\beta} \} = 2i\,{\cal
D}_{\alpha \dot\beta} - 2i\,({\cal D}_{\alpha}\psi^\gamma {\overline{\cal
D}}_{\dot\beta}  {\bar\psi}^{\dot\gamma})\, {\cal D}_{\gamma\dot\gamma}~, \nn
&&\{{\cal D}_\alpha, \,{\cal D}_{\gamma\dot\gamma}\} = -2i\,( {\cal
D}_{\alpha}\psi^\beta {\cal D}_{\gamma\dot\gamma} {\bar\psi}^{\dot\beta})\,
{\cal D}_{\beta\dot\beta}~. \label{cd4alg} \eea
The first two relations in \p{cd4alg} guarantee the consistency of the above
nonlinear version of $N=1, \;d=4$ chirality. They also imply, like in the flat
case,
\be
({\cal D})^3 = ({\overline{\cal D}})^3 = 0~. \label{cubvan}
\ee

The second flat irreducibility constraint, eq.
\p{constr5}, is not so simple to covariantize. The straightforward
generalization of \p{constr5},
\be\label{constr5f}
{\cal D}^{\alpha}\psi_{\alpha}+
 {\overline{\cal D}}_{\dot\alpha}{\overline \psi}^{\dot\alpha}=0~,
\ee
is contradictory. Let us apply the square $({\cal D})^2$ to
the  left-hand side of \p{constr5f}. When hitting the first term in the sum,
it yields zero in virtue of the property \p{cubvan}. However, it
is not zero on the second term. To compensate
for the resulting non-vanishing terms, and thus to achieve compatibility
with the algebra \p{cd4alg} and its corollaries \p{cubvan}, one should
modify \p{constr5f} by some  higher-order coorrections
\cite{BG2}.

Let us argue that the constraints \p{constr5} {\it together} with the
equations of motion \p{eom5} can be straightforwardly covariantized as
\be\label{constr5h} {\cal D}^{\alpha}\psi_{\alpha}=0\;,\quad  {\overline{\cal
D}}_{\dot\alpha}{\overline \psi}^{\dot\alpha}=0 \;.
\ee

Firstly, we note that no difficulties of the above kind related to the
compatibility with the  algebra \p{cd4alg} arise on the shell of eqs.
\p{constr5h}. As a consequence of \p{constr5h} and the first two relations
in \p{cd4alg}  we get
\be
{\cal D}^2\,\psi_{\alpha}=0,\quad {\overline{\cal
D}}^2\,\bar\psi_{\dot\alpha}=0\;. \label{squarenonl}
\ee
This set is a nonlinear version of the well-known reality condition and
the equation of motion for the auxiliary field of vector
multiplet. Then, applying, e.g.,
${\cal D}_\alpha$ to the second equation in \p{constr5h} and making use of the
chirality condition \p{chir5e}, we obtain the nonlinear version of the
equation of motion for photino
\be
{\cal D}_{\alpha \dot\alpha}\bar\psi^{\dot\alpha} - ({\cal
D}_{\alpha}\psi^\gamma {\overline{\cal D}}_{\dot\alpha}
{\bar\psi}^{\dot\gamma})\, {\cal D}_{\gamma\dot\gamma}\bar\psi^{\dot\alpha} =
0~.
\ee
Acting on this equation by one more ${\cal D}_\alpha$ and taking advantage of
the equations \p{squarenonl}, we obtain:
\be
\left[{\cal D}^\alpha, {\cal D}_{\alpha{\dot\alpha}}
\right]\bar\psi^{\dot\alpha}- {\cal D}_{\alpha}\psi^\gamma
 \{ {\cal D}^\alpha ,{\overline{\cal D}}_{\dot\alpha}\}
\bar\psi^{\dot\gamma} {\cal D}_{\gamma\dot\gamma}
\bar\psi^{\dot\alpha} -
{\cal
D}_{\alpha}\psi^\gamma{\overline{\cal
D}}_{\dot\alpha}\bar\psi^{\dot\gamma} \left[
{\cal D}^\alpha , {\cal D}_{\gamma\dot\gamma}\right] \bar\psi^{\dot\alpha}
=0~. \label{secstep}
\ee
After substituting the explicit expressions for the (anti)commutators from
\p{cd4alg}, we observe that \p{secstep} is satisfied identically, i.e. it
does not imply any further restrictions on $\psi^\alpha,
\bar\psi^{\dot\alpha}$. It can be also explicitly checked, in a few lowest
orders in $\psi^\alpha, \bar\psi^{\dot\alpha}$, that the higher-order
corrections to \p{constr5f} found in \cite{BG2} are vanishing on the
shell of eqs. \p{constr5h}.

Thus the full set of equations describing the dynamics of the D3-brane
supposedly consists of the generalized chirality constraint \p{chir5e}
and the equations \p{constr5h}. To prove its equivalence to the $N=1$
superfield description of D3-brane proposed in \cite{BG2}, recall that
the latter is the $N=1$ supersymmetrization \cite{CF} of the $d=4$ Born-Infeld
action  with one extra nonlinearly realized $N=1$ supersymmetry. So, let us
consider the bosonic part of the proposed set of equations.
Our superfields
$\psi,\bar\psi$ contain the following bosonic components: \be\label{defcomp5}
V^{\alpha\beta}=   V^{\beta\alpha}\equiv {\cal
D}^{\alpha}\psi^{\beta}|_{\theta=0}~, \quad {\bar V}^{{\dot\alpha}\dot\beta}=
{\bar V}^{{\dot\beta}\dot\alpha}\equiv   {\overline{\cal
D}}^{\dot\alpha}{\bar\psi}^{\dot\beta}|_{\theta=0}\;, \;
\ee
which, owing to \p{constr5h}, obey the following simple equations
\be\label{eom6}
\partial_{\alpha\dot\alpha}V^{\alpha\beta}-V_{\alpha}^{\gamma}
 {\bar V}_{\dot\alpha}^{\dot\gamma}\;\partial_{\gamma\dot\gamma}
  V^{\alpha\beta} =0~,\quad
\partial_{\alpha\dot\alpha}{\bar
V}^{{\dot\alpha}\dot\beta}-V_{\alpha}^{\gamma}
 {\bar V}_{\dot\alpha}^{\dot\gamma}\;\partial_{\gamma\dot\gamma}
  {\bar V}^{{\dot\alpha}\dot\beta} =0 \;.
\ee
Like in the D2-brane case, in the equations \p{eom6} nothing
reminds us of the Born-Infeld equations. Nevertheless, it is possible to
rewrite these equations in the standard Born-Infeld form.

The first step is to rewrite eqs.\p{eom6} as
\bea
&&\left(1-\frac{1}{4} V^2{\bar V}{}^2\right)\partial_{\beta\dot\alpha}
V^{\beta}_{\alpha}+ \frac{1}{4} {\bar V}{}^2 V_{\alpha}^{\beta}
 \partial_{\beta\dot\alpha}V^2+
  \frac{1}{2}{\bar
V}{}_{\dot\alpha}^{\dot\beta}\partial_{\alpha\dot\beta}
  V^2 =0 \;, \label{1}\\
&&\left(1-\frac{1}{4} V^2{\bar V}{}^2\right)\partial_{\alpha\dot\beta}
{\bar V}{}^{\dot\beta}_{\dot\alpha}+
 \frac{1}{4} V^2 {\bar V}{}_{\dot\alpha}^{\dot\beta}
 \partial_{\alpha\dot\beta}{\bar V}{}^2+
  \frac{1}{2} V_{\alpha}^{\beta}\partial_{\dot\alpha\beta}
  {\bar V}{}^2 =0 \;.\label{2}
\eea
After some algebra, one can bring them into the following equivalent form
\be\label{last1}
\partial_{\beta\dot\alpha}\left(fV_{\alpha}^{\beta}\right) -
  \partial_{\alpha\dot\beta}\left( {\bar f}{\bar
V}_{\dot\alpha}^{\dot\beta}
  \right) =0~, \quad
\partial_{\beta\dot\alpha}\left(gV_{\alpha}^{\beta}\right) +
  \partial_{\alpha\dot\beta}\left( {\bar g}{\bar
V}_{\dot\alpha}^{\dot\beta}
  \right) =0 \;,
\ee
where
\be\label{sol1}
f=\frac{ {\bar V}{}^2-2}{1-\frac{1}{4}V^2{\bar V}{}^2}~, \quad
g=\frac{ {\bar V}{}^2+2}{1-\frac{1}{4}V^2{\bar V}{}^2}\; .
\ee
After introducing the ``genuine'' field strengths
\be\label{fc5}
F_\alpha^\beta\equiv \frac{1}{2\sqrt{2}}\, f\,V_\alpha^\beta~, \quad
{\bar F}_{\dot\alpha}^{\dot\beta}\equiv \frac{1}{2\sqrt{2}}\,
  {\bar f}\,{\bar V}_{\dot\alpha}^{\dot\beta}~,
\ee
first of eqs. \p{last1} is recognized as the Bianchi
identity
\be
\partial_{\beta\dot\alpha}F_{\alpha}^{\beta} -
  \partial_{\alpha\dot\beta}{\bar F}_{\dot\alpha}^{\dot\beta}   =0 \;,
\ee
while the second one acquires the familiar form of the Born-Infeld equation
\bea\label{last2} &&\partial_{\beta\dot\alpha}\left(
\frac{1+F^2-{\bar F}{}^2}{\sqrt{(F^2-{\bar F}{}^2)^2-2(F^2+{\bar
F}{}^2)+1}}
F_{\alpha}^{\beta}\right) \nn
&& +\; \partial_{\alpha\dot\beta}\left(
\frac{1-F^2+{\bar F}{}^2}{\sqrt{(F^2-{\bar F}{}^2)^2-2(F^2+{\bar
F}{}^2)+1}}
{\bar F}_{\dot\alpha}^{\dot\beta}
  \right) =0 \;.
\eea
Thus, in this new basis the action for our bosonic system is
the Born-Infeld action:
\be
S=\int d^4x \sqrt{(F^2-{\bar F}{}^2)^2-2(F^2+{\bar F}{}^2)+1} \;.
\ee

Now the equivalence of the system \p{constr5h} to the equations corresponding
to the action of ref. \cite{BG2}, like in the D2-brane case, can
be established proceeding from the following two arguments: (i) It is
$N=1$ supersymmetrization of the $d=4$ Born-Infeld  equations; (ii) It
possesses the second hidden nonlinearly realized  supersymmetry lifting $N=1,
\, d=4$ to $N=2, \,d=4$. The action given in
\cite{BG2} provides the unique extension of the $d=4$ Born-Infeld action
with both these requirements satisfied. Hence, both representations should be
equivalent to each other.

Note that at the full superfield level the redefinition \p{fc5} should
correspond to passing from the Goldstone fermions  $\psi_\alpha$,
$\bar\psi_{\dot\alpha}$  which have the simple transformation properties
in the nonlinear realization of $N=1, \,d=4$ supersymmetry but obey the
nonlinear irreducibility constraints, to the ordinary Maxwell superfield
strength $W_\alpha, \,\bar W_{\dot\alpha}$ defined by eqs. \p{chir5},
\p{constr5}. The nonlinear action in \cite{BG2} was written just in terms of
this latter  object. The equivalent form \p{constr5h} of the equations of
motion and  Bianchi identity is advantageous in that it is manifestly
covariant under  the second (hidden) supersymmetry, being constructed out of
the  covariant objects.

\vspace{0.3cm}

\noindent{\bf 4. Conclusions.} In this Letter we demonstrated that in
many cases one can simplify the analysis of the equations of motion which
follow from the coset approach by taking no account of the automorphism
group at all. We showed that the equations of motion
for the $N=1,\,D=4$ supermembrane, D2- and D3-branes in a flat
background have a very simple form when written in terms of Goldstone
superfields of nonlinear realizations and the corresponding nonlinear covariant
derivatives. As a by-product, we got a new simple form for the $d=3$ and $d=4$
Born-Infeld theory equations of motion combined with the appropriate Bianchi
identities. The remarkable property of this representation is that it
involves  only a third order nonlinearity in the gauge field
strength.

Note that the idea to use the geometric and symmetry principles to derive the
dynamical equations is not new, of course. For instance, the completely
integrable $d=2$ equations admit the geometrical interpretation as the
vanishing of some curvatures. In the superembedding approach (see \cite{dima}
and refs. therein) the equations of motion for superbranes in a number of
important cases amount to the so-called ``geometro-dynamical'' constraint
which, in the PBGS language, is just a kind of the inverse Higgs constraints.
For instance, this applies to the $N=1, \,D=10$ 5-brane \cite{BIK1,BIK2}. In
this case the condition like \p{basconstr}, besides
eliminating the Goldstone fermion  superfield in terms of the appropriate
analog of the $d=3$ superfield $q$ ($d=6$ hypermultiplet superfield), also
yields the equation of motion for the latter \footnote{It is curious that this
equation, in accord with the general reasoning of the present work, proved
to be finally written in terms
of the covariant quantities of nonlinear realization of the pure $N=1, \,D=10$
Poincar\'e superalgebra, despite the fact that we started in \cite{BIK1,BIK2}
from the coset of the extended supergroup involving $D=10$ Lorentz group.}.
However, as we saw  in the above examples, in other interesting cases the
inverse Higgs (or geometro-dynamical) constraints do not imply any
dynamics which, however, can still be implemented in a manifestly covariant
way using the approach proposed here.

It still remains to fully understand why in the PBGS scheme the dynamical
worldvolume superfield equations are not sensitive to the presence or absence
of the automorphism  generators in the initial coset construction. This is in
contrast with the case of purely bosonic $p$-branes. For the self-consistent
description of them in terms of nonlinear realizations one should necessarily
make use of the cosets of the full target Minkowski space Poincar\'e group
including the Lorentz (automorphism) part of the latter \cite{PW,EI}. A
possible explanation of this apparent disagreement is that the
Goldstone fermion superfields or  Goldstone superfields associated with the
central charges (and/or with the transverse components of the full momenta)
already accommodate the Lorentz and other automorphism groups Goldstone fields.
These come out as component fields in the $\theta$ -expansion of the
Goldstone superfields. So the automorphism groups Goldstone fields are
implicitly present in  the superbrane superfield equations of motion.

The most interesting practical application of the approach
exemplified here is the possibility to construct, more or less
straightforwardly, the equations for the $N=4$ and $N=8$ supersymmetric
Born-Infeld theory.This work is in progress now \cite{BIK33}.

\vspace{0.3cm}

\noindent{\bf Acknowledgements.}\hskip 1em
This work was supported in part by the Fondo Affari Internazionali
Convenzione Particellare INFN-JINR, grants RFBR-CNRS 98-02-22034, RFBR
99-02-18417, INTAS-96-0538, INTAS-96-0308 and NATO Grant PST.CLG 974874.

\end{document}